**TÍTULO: Visões da Indústria 4.0**
**CRISTINA DIAS SANT' ANNA, WALLACE CAMACHO CARLOS**

**RESUMO:**

A indústria é a parte de uma economia que produz bens materiais altamente mecanizados e automatizados. Desde o início da industrialização, houveram várias etapas e mudanças de paradigmas que hoje são ex-pós-chamadas de "revoluções industriais": no campo da mecanização (chamada 1ª revolução industrial), do uso intensivo de energia elétrica (chamada 2ª revolução industrial) e da digitalização generalizada (chamada 3ª revolução industrial). Nesse sentido, por essa expectativa futura, o termo "Indústria 4.0" foi estabelecido para uma "4ª revolução industrial".

A evolução especialmente na Europa, mas também nos Estados Unidos, cunhado como Internet Industrial é frequentemente comparada com a continuação de aumentos disruptivos na produção industrial, como as revoluções iniciadas por vapor, eletricidade e etc.

Aspectos em treinamento contínuo da força de trabalho, e a utilização de recursos de sustentabilidade em contextos industriais, econômicos e políticas gerais de governança em TI não estão amplamente difundidas e são os principais problemas e desafios em paradigmas na Indústria 4.0. Direções das pesquisas temáticas futuras que serão abordados neste artigo.

**Palavras-chave:** Indústria 4.0; Global; Revolução Industrial; Mudanças Industriais.

**TITLE: Industry 4.0**
**CRISTINA DIAS SANT' ANNA, WALLACE CAMACHO CARLOS**

**ABSTRACT:**

Industry is part of an economy that produces highly mechanized and automated material goods. Since the beginning of industrialization, there have been several stages and paradigm shifts that today are ex-post-so-called "industrial revolutions": in the field of mechanization (called the 1st industrial revolution), the intensive use of electrical energy (called the 2nd industrial revolution) and widespread digitization (called the 3rd industrial revolution). In this sense, for this future expectation, the term "Industry 4.0" was established for a "4th industrial revolution".

Developments especially in Europe, but also in the United States, coined as the Industrial Internet, are often compared with the continuation of disruptive increases in industrial production, such as revolutions initiated by steam, electricity, etc.

Aspects of continuous workforce training, and the use of sustainability resources in industrial, economic and general IT governance policies are not widespread and are the main problems and challenges in paradigms in Industry 4.0. Directions for future thematic research that will be covered in this article.

1. **INTRODUÇÃO**

A evolução de novas tecnologias no mercado industrial em face das novas tendências das redes de comunicação e plataformas de dados estamos diante de um cenário que vai nos permitir implementar a chamada quarta "Revolução Industrial", Indústria 4.0.

As estratégias de negócios adotadas no uso de tecnologias devem olhar para padrões abertos de mercado, arquiteturas de referências, treinamento e desenvolvimento de profissionais. Por um lado, há uma enorme atração de aplicativos, que induz uma necessidade notável de mudanças devido às mudanças nas condições do quadro operacional. Os gatilhos para isso são mudanças sociais no contexto dos aspectos ecológicos, e que requerem um foco mais intenso na sustentabilidade em contextos industriais, econômicos e políticas gerais. Por outro lado, há uma demanda tecnológica na prática industrial. No entanto, em relação ao trabalho, especialmente nos contextos industriais, as tecnologias inovadoras não estão amplamente difundidas. Portanto, uma descrição das tecnologias utilizadas e as qualificações no trabalho são citadas neste artigo.

Esta é uma área extremamente desafiadora, uma vez que normas e padrões devem ser aplicados não apenas em diferentes países, mas também em diferentes sistemas. Além disso, a natureza altamente dinâmica da tecnologia exige que eles sejam altamente flexíveis e capazes de se adaptarem. Idealmente, os padrões ou normas devem ser estabelecidos para soluções internacionais ou corporativas, a fim de criar um investimento seguro no ambiente de desenvolvimento além de construir confiança. (KAGERMANN et al., 2016).

Sendo assim, o objetivo geral deste artigo pretende descrever a evolução e os principais paradigmas e direções de tendências de tecnologia da Indústria 4.0. Cinco tecnologias desenvolvidas em digitalização são identificadas e considerações finais em indústria 4.0 são discutidas no final deste documento.

## 2. METODOLOGIA

Quanto à forma de abordagem do problema e aos seus objetivos, esta pesquisa classifica-se como descritiva, pois envolve técnicas padronizadas de coleta de dados e visa descrever as características de determinada população ou fenômeno (Silva e Menezes, 2005).

Nesse sentido, será feito levantamentos bibliográficos e estudos de casos na pesquisa exploratória sobre Indústria 4.0 e dentre outros levantamentos bibliográficos sobre a produtividade em novos paradigmas de cenários tecnológicos. (Silva e Menezes, 2005).

## 3. REVISÃO BIBLIOGRÁFICA

### 1. O CAMINHO DA EVOLUÇÃO

Combinando os pontos fortes da fabricação industrial otimizada com tecnologia de ponta, as tecnologias da Internet são o núcleo da Indústria 4.0. Portanto, não é surpresa que a indústria 4.0 esteja recebendo uma atenção cada vez mais crescente, especialmente na Europa, mas também nos Estados Unidos, cunhado como Internet Industrial. Indústria 4.0 é frequentemente comparada com a continuação de aumentos disruptivos na produção industrial, como as revoluções iniciadas por vapor, eletricidade e etc. Semelhante à Indústria 4.0, estas "Revoluções" foram iniciadas não por uma única tecnologia, mas pela interação de número de avanços tecnológicos cujos efeitos quantitativos criaram novas formas de produção (Schmidt et al., 2015).

Em geral, o desenvolvimento da Indústria 4.0 significa o aparecimento de "fábricas inteligentes", onde as máquinas são capazes de se adaptar às mudanças reconfigurando-se e otimizando-se novamente, e onde os dados de vários recursos são alinhados aos processos digitalizados. (Ekaterina Uglovskaia, 2017). As constantes evoluções tecnológicas e também os paradigmas como as empresas estão lidando com seus dados sendo gerados e capturados a medida que clientes interagem com seus produtos tem exigido a necessidade de uma maior compreensão com processos analíticos e integrados com soluções de tecnologia escalável e de alto desempenho para integrar componentes e processos evolutivos e adaptáveis à medida que exigem novas atualizações e melhorias contínuas.

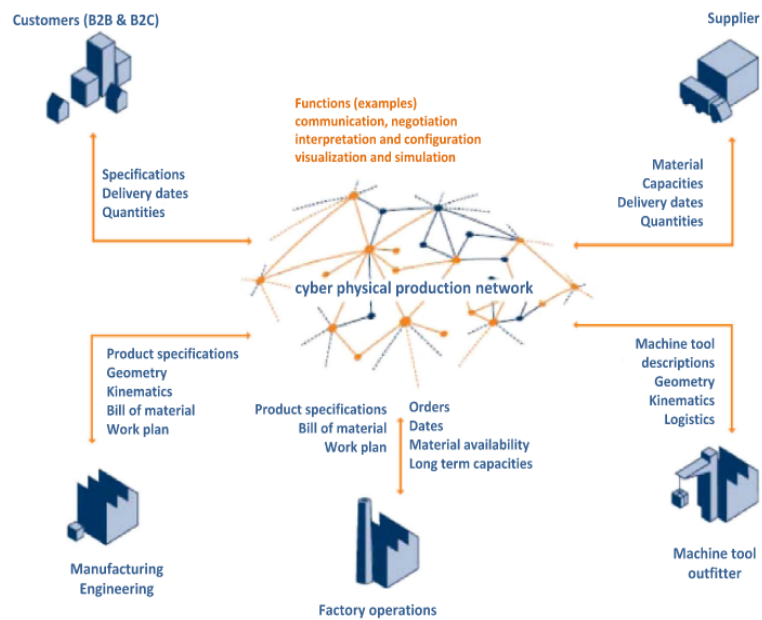

Figura 1 - Exemplo de interdependências de uma cadeia de abastecimento no contexto do futuro projeto "Indústria 4.0". Fonte: Geisberger and Broy (2012, p.56).

Para o cenário proposto no exemplo da figura 1 em termos "Sistema Físico Cibernético" - (CPS), que são sistemas embarcados inteligentes caracterizados por hardware e software e operam tanto em nível virtual quanto físico, interagindo e controlando dispositivos físicos, detectando e mudando o mundo real. (Baheti, 2011). Este cenário abrange uma visão de integração entre componentes e cenários ligados a fábricas e cidades inteligentes que se comunicam através de uma malha de internet das coisas, destruídas e alimentadas por redes de internet de alta conexão.

## 2. TECNOLOGIAS DESENVOLVIDAS NA INDÚSTRIA 4.0 - DIGITALIZAÇÃO

### 1. Digitalização de ativos físicos - Digital Twins

Empresas têm investido em modelos de digitalização de ativos como fábricas, componentes, processos e linhas de montagem. Isso permite a previsão e manutenção prescritiva além de permitir desenhar e visualizar de modo interativo processos realizados por máquinas e também processos por ações humanas dentro de uma tarefa mecanizada dentro de um processo fabril. Para isso requer uma plataforma com grandes transferências de dados em tempo real, volumes de dados e processamento de alto desempenho para estas transformações.

**2. Manutenção Preditiva**

Sensores IIoT são usados para coletar dados e um sistema carregado com um modelo de Aprendizado de Máquina (ML) implantado em uma nuvem pública para analisar anomalias de dados. Isso permite que a manutenção das máquinas seja realizada de forma proativa, diminuindo o tempo de inatividade e melhorando a capacidade de identificar a obsolescência de componentes e eventuais desgastes mecânicos de peças que exigem uma utilização constante ficando expostas a maiores desgastes físicos.

**3. Serviços de localização em tempo real - (RTLS) Real Time Location Services**

Sistemas de rastreamento e monitoramento aumentam a segurança dos funcionários e protegem os ativos do fabricante. Usar RTLS para evitar roubo pode economizar para grandes empresas uma significativa fatia do seu orçamento anual com gastos para reposição de perdas e furtos de equipamentos. Esses dispositivos podem ter componentes IIOS (Industrial Internet of Things), conectados a recursos em nuvem e monitoramento em tempo real de ativos. Esses sistemas podem responder a eventos e notificarem pessoas ou outros sistemas, respondendo a configurações ou métricas estipuladas. Variações e limites de localidades podem ser identificados dentro de um raio de atuação de um segmento e também permitir facilidades de gerenciamento na mobilidade de recursos e automatização de robôs, a empregabilidade das tarefas de monitoramento podem estar empregadas em sistemas que calculam o momento exato da chegada de uma matéria prima destacando para sua utilização e recepção do insumo.

**4. Realidade Aumentada - RA**

A utilização de interfaces ricas e inteligentes que se projetam dentro de dispositivos celulares e tablets, refletido a partir de ambientes físicos reais, tem vislumbrado utilização deste segmento tecnológico muitas das vezes voltado para a indústria de games como um novo modelo e plataforma para o mundo real com empregabilidade em simulações de equipamentos, treinamento de operações, detalhamento de peças e instruções e manutenção de equipamentos com assistência remota fazendo uso da tecnologia eliminando dificuldades e barreiras de idiomas. Assim, informações de manutenção em campo que muitas vezes são de difícil interpretação e requerem a experiência do operador podem ser simuladas nos telemóveis ou tablets, reduzindo os custos de deslocamentos, evitando interpretações erradas e consequentemente retrabalhos nas ações de manutenção.(B. P. Santos 1* , A. Alberto 2 , T.D.F.M.Lima 1 , F.M.B. Charrua-Santos 1). A utilização de dispositivos como óculos inteligentes. A microsoft criou um dispositivo chamado Hololens - óculos de realidade imersiva com recursos de voz, som e imagem, que se comunicam por interfaces de SDK (Software Development Kit) disponíveis para desenvolvimento para a sua plataforma de realidade imersiva, com recursos de inteligência prontos para criação de softwares personalizados.

## 5. Redes 5G na indústria 4.0

Em um cenário projetado para pelo menos os próximos três ou quatro anos, a evolução das redes celulares alcançará mais um grande salto em termos de evolução na cadeia de comunicação e dados, com a utilização e implantação dos padrões de tecnologias de redes celulares 5G. Esta tecnologia será mais um passo para a comunicação e a utilização de banda larga sem fio. As redes 5G prometem aos seus futuros usuários uma cobertura mais ampla e eficiente, maiores transferências de dados, além de um número significativamente maior de conexões simultâneas.

Os critérios técnicos que podem ser destacados como uma vantagem na utilização da tecnologia de quinta geração: as redes 5G devem consumir até 90% menos energia que as redes 4G e os modelos anteriores; os tempos de conexão e taxas de latência entre aparelhos móveis devem ser inferiores a 5 ms (milissegundos), em relação à latência de 30 ms das redes 4G; o número de aparelhos conectados por área devem ser 50 a 100 vezes maior que o atual. Nesse cenário, a comunicação se move além da comunicação entre pessoas, incluindo a comunicação entre máquinas, o que gera um impacto fundamental na arquitetura da rede. Antigos problemas se tornam possíveis para serem resolvidos e soluções antigas são substituídas por novas. Surgirão novas oportunidades de negócios e reformulação dos processos industriais existentes(Alex Vidibal Basto).

Com a implantação das redes 5G, e os seus benefícios para utilização de conexões com baixas latências, diversas possibilidades e novas arquiteturas de sistemas são empregadas para extrair o máximo de recursos que esta tecnologia pode oferecer às indústrias. O uso da comunicação D2D (Dispositivo Para Dispositivo) em redes celulares é definida como uma comunicação direta entre dois dispositivos móveis sem passar pela estação base (BS) ou núcleo da rede. Comunicação D2D não é transparente para a rede celular e pode ocorrer dentro (Inband) ou fora (Outband) do espectro celular (Alex Vidibal Basto). Este modelo de comunicação pode ser utilizado na comunicação de máquinas com outras máquinas e equipamentos integrados com componentes IIOT. Os dados gerados entre as máquinas podem ser enviados através da própria rede 5G para serviços de nuvem pública, realizando integrações entre sistemas de monitoramento e detecção de falhas mecânicas.

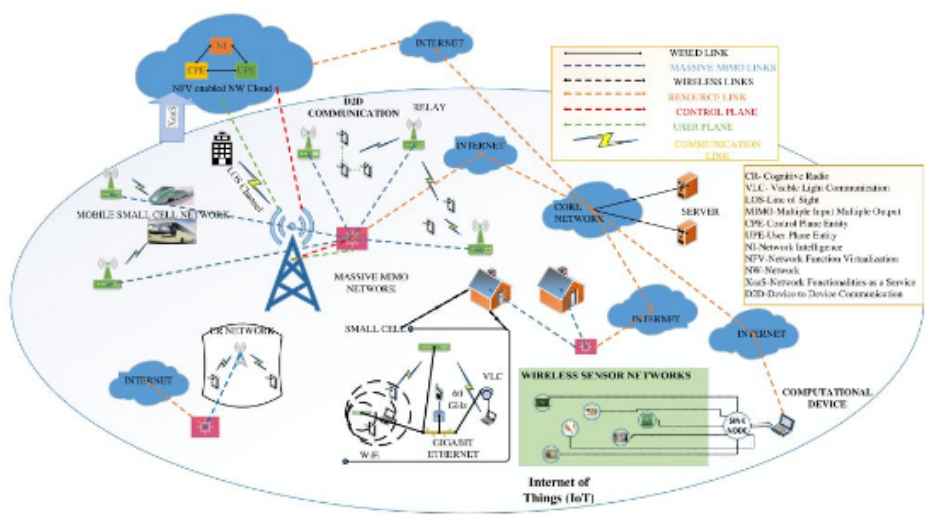

Figura 2 - Arquitetura Geral de 5G de redes celulares. Fonte: Gupta and Jha (2015).

## 3. QUALIFICAÇÕES E HABILIDADES

No que diz respeito à força de trabalho, as fábricas inteligentes modernas exigirão tecnicamente funcionários especializados e qualificados. Devido à alta automação de processos, a tendência será cortar postos de trabalho, enquanto o restante dos empregos na indústria incluirá mais tarefas de curto prazo, mas difíceis de planejar. Qualificações dos trabalhadores incluirá conhecimento e proficiência de TI, processamento e análise de dados, conhecimento estatístico e compreensão organizacional, capacidade de adaptação e interagir usando interfaces homem-máquina modernas (Gehrke, 2015). Será sempre um esforço maior de capacitação e treinamento contínuo de colaboradores e programas de incentivo na iniciativa privada e também pública para alavancar áreas de conhecimento restrito dos segmentos de pesquisa. Formação de base em escolas técnicas com ênfase em computação, engenharia, matemática e estatística como requisitos técnicos são amplamente valorizados e essenciais para formação básica de profissionais qualificados a operar em ambientes colaborativos e inovadores.

Por isso, é esperado um apoio maior de áreas tecnológicas que sustentam o conhecimento nesta base da pirâmide em figura 3. Ferramentas tecnológicas e processos organizacionais.

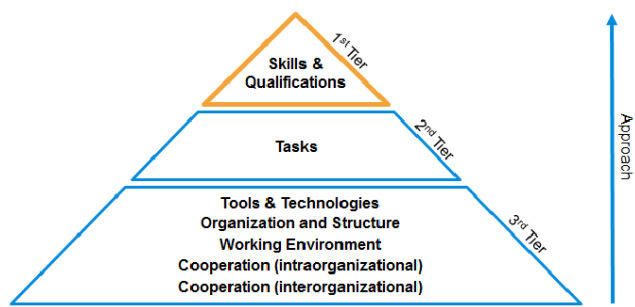

Figura 3 - Abordagem para derivar qualificações e habilidades para o operário do futuro. Fonte: Gehrke (2015).

## 4. PRINCÍPIOS DE DESIGN E TENDÊNCIAS DE TECNOLOGIA NA INDÚSTRIA 4.0

Para garantir precisão e confiabilidade da análise de conteúdo, e para entender melhor como a academia define a Indústria 4.0 com base em seus princípios de design e tendências de tecnologia, apenas artigos de periódicos e livros/seções de livros publicados por editoras acadêmicas de renome (incluindo Elsevier, Springer, Emerald, Acatech) foram selecionados. A Indústria 4.0 é um sistema integrativo de criação de valor que é composto por 12 princípios de design e 14 tendências de tecnologia.

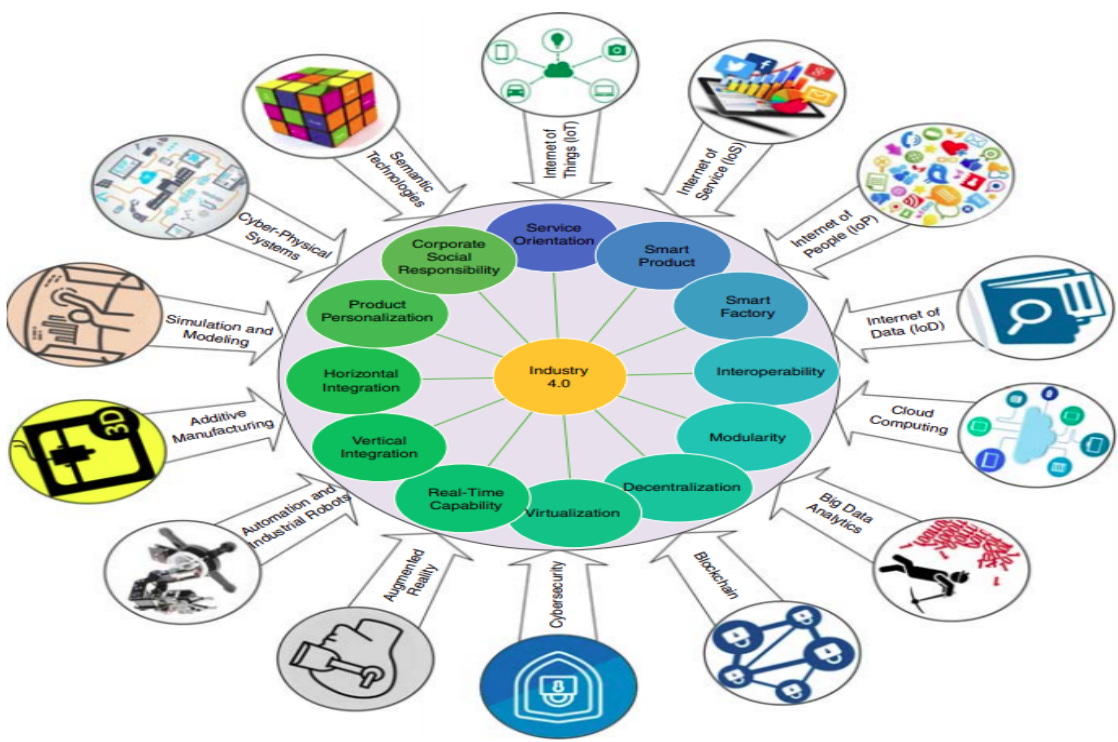

Figura 4 - Princípios de design e tendências de tecnologia. Fonte: Ghobakhloo Morteza (2018).

A figura 4 identifica a parte do círculo interno que representa os princípios de design e o círculo externo inclui as tendências de tecnologia na Indústria 4.0.

A computação em nuvem não é um conceito completamente novo, mas não existe um conceito universal ou definição padrão de computação em nuvem. Este paradigma evoluiu com base nos avanços recentes em hardware, tecnologia de virtualização, computação distribuída e entrega de serviços na internet (Oliveira et al., 2014).

A aplicação de computação em nuvem oferece aos fabricantes com aplicativo de software baseado em nuvem, painel de gerenciamento baseado na web e baseado em nuvem colaborativa e permite a integração de recursos de manufatura distribuídos e estabelecimento de uma infraestrutura colaborativa e flexível distribuída geograficamente em locais de manufatura e serviços (He e Xu, 2015). Isso, por sua vez, levará à nuvem manufatura como o paradigma de manufatura de próxima geração (Ooi et al., 2018).

Blockchain, também conhecido como tecnologia de razão distribuída, é a base da criptomoedas como Bitcoin e Ethereum, mas seus recursos vão muito além. A comunidade científica acredita que a tecnologia blockchain é crítica para a Indústria 4.0 porque criptomoedas permitem que incontáveis dispositivos inteligentes funcionem de forma transparente, segura, rápida e transações financeiras sem atrito, totalmente autônomas, sem intervenção humana no ambiente IoT (Devezas e Sarygulov, 2017; Sikorski et al., 2017). Por exemplo, colocando blockchain entre IIoT, sistemas de produção ciberfísicos e parceiros de fornecimento podem permitir que máquinas dentro da fábrica inteligente façam um pedido de forma segura e autônoma para suas peças de reposição para otimizar ainda mais os processos e prometem inúmeros benefícios, como tempo de ciclo de peça reduzido, menor taxa de defeitos, maior qualidade e confiabilidade, redução do desperdício e melhor utilização do espaço físico, tornando-se indispensável para fabricantes de classe mundial (Esmaeilian et al., 2016).

A segurança cibernética é um elemento-chave da Indústria 4.0, uma vez que todas as organizações com internet estão em risco de ataque. O Stuxnet nunca pode ser esquecido, o notório malware que infectou sistemas de controle nas usinas nucleares e manipulava a velocidade das centrífugas, causando eles girarem fora de controle. Não há dúvida de que a Indústria 4.0 será desafiada por questões tradicionais de cibersegurança, juntamente com seus próprios problemas exclusivos de segurança e privacidade (Thames e Schaefer, 2017).

Técnicas de simulação e modelagem visam a simplificação e o favorecimento econômico de projetar, realizar, testar e executar uma operação ao vivo de sistemas de manufatura (Kocian et al., 2012). Por exemplo, os fabricantes hoje em dia podem simular a usinagem de peças usando dados da máquina física que levam a redução do tempo de configuração para o processo de usinagem real em até 80 por cento (Rüßmann et al., 2015). Relatórios industriais revelam que os fabricantes de classe mundial visualizaram o maior potencial de simulação no futuro em uma tentativa de teste virtual de sistemas de produção.

De perspectiva de sustentabilidade ambiental, a Indústria 4.0 oferece imensas oportunidades para o realização de manufatura sustentável, pois permite a coordenação eficiente do produto, material e energia em todos os ciclos de vida do produto; design sustentável de produtos; desenho sustentável de processos e materialização da eficiência de recursos; uma maior eficiência dos trabalhadores graças à infraestrutura IIoT; e implantação do modelo de negócios greenleagile (Stock e Seliger, 2016).

Indústria 4.0 é um sistema dinâmico e integrado para exercer controle sobre toda a cadeia de valor do ciclo de vida dos produtos. Integração vertical e horizontal e fusão do físico e do mundos virtuais estão no centro da Indústria 4.0 e tendências de tecnologia como CPS, IIoT, IoS e IoP, Blockchain e WoT permitem esse nível de integração em escala global.

A transição digital exigida pela Indústria 4.0 não só desafia a capacidade das empresas de inovar, mas também requer novas estratégias e modelos organizacionais, e mudanças em toda a organização em infraestrutura física, operações e tecnologias de manufatura, recursos humanos e gestão das práticas (Gilchrist, 2016).

Estratégias devem ser definidas em um plano baseado no tempo e descrever onde a empresa está, onde precisa ir e como chegar lá, com base nas visões pré-definidas e planos na Indústria 4.0 (Schumacher et al., 2016).

Não são todas as organizações que têm maturidade de TI adequada para adotar a Indústria 4.0, e nem todos fabricantes com produção habilitada para IoT ou sistemas de serviço são grandes o suficiente para lidar com a integração horizontal e sustentar sua posição competitiva no mundo globalizado e mercado hipercompetitivo (Gilchrist, 2016; Leyh et al., 2016). Isso significa que a Indústria 4.0 necessita de inúmeras aquisições e fusões (M & A) em escala global, e fabricantes de qualquer tamanho que objetivam a digitalização devem planejar cuidadosamente potenciais fusões e aquisições com antecedência.

Os princípios de design e tendências de tecnologia da Indústria 4.0, como horizontal e integração vertical, IIoT, IoD, CPS, interoperabilidade, simulação e blockchain indicam que a quarta revolução industrial gira em torno de TI. A governança de TI é normalmente o mais fraco aspecto de governança corporativa (Wu et al., 2015), e como a primeira etapa da estratégia de maturidade de TI para a Indústria 4.0, os fabricantes devem garantir que a equipe de governança de TI esteja pronta para elaborar estratégias, orçamentos, executar, controlar e relatar projetos e operações de avanço de TI de acordo com os requisitos de transição da Indústria 4.0. A equipe de governança de TI deve realizar uma análise detalhada da infraestrutura de TI existente (por exemplo, redes, hardware de computador e software, sensores, controladores e atuadores) e identificar a abordagem mais significativa para usá-los no suporte à transição da Indústria 4.0. A equipe de governança de TI deve ainda identificar diferentes segmentos de negócios que precisam de rede e integração, e quando a infraestrutura de TI existente não suportar totalmente a digitalização desses segmentos de negócios, formular e implementar as estratégias de desenvolvimento de TI necessárias (Savtschenko et al., 2017).

## 4. CONSIDERAÇÕES FINAIS

A evolução tecnológica discutida retrata um cenário de uma nova possibilidade de revoluções nas indústrias assim como nas empresas que usam recursos de T.I de forma inteligente para alinhar estratégias táticas e operacionais com o máximo de eficiência apoiada por ferramentas de tecnologia. Mesmo assim, é importante que as empresas se lembrem de que a adoção do conceito da Indústria 4.0 não envolve apenas a tecnologia mais recente, deve começar com as raízes, o que inclui as pessoas, a cultura, a estratégia corporativa e os processos de governança de TI. Além disso, uma investigação cuidadosa das possibilidades, benefícios e riscos de segurança cibernética devem ser realizados. Uma das premissas para se alcançar o conhecimento adequado para dar o primeiro passo na modernização dos processos e também de componentes que integram as corporações sempre será uma mão de obra qualificada e técnica para compor segmentos onde são necessárias habilidades e qualificações para um direcionamento de sucesso na utilização dos meios existentes da tecnologia. O conjunto de habilidades mencionadas, que é recomendado para a força de trabalho do futuro, consiste em conhecimento técnico e habilidades pessoais. Em um cenário de constante evolução, as corporações e indústrias devem realizar análises constantes e avaliar seu grau de maturidade e absorção de novas tecnologias pois o caminho para alcançar o estado da arte na utilização de novas tecnologias não acontecem da noite para o dia e devem ser seguidas de evoluções constantes e pequenas para atingir os seus objetivos alinhados com as estratégicas de negócio.

Como mencionado, não são todas as organizações que têm maturidade de TI adequada para adotar a Indústria 4.0. Isso significa que a Indústria 4.0 necessita realizar um planejamento cuidadoso dos recursos tecnológicos em potencial com antecedência.

# 5. REFERÊNCIAS